\documentclass[10pt,a4paper,twoside]{article}
\pdfoutput=1
\usepackage{epsfig}
\usepackage{baltlat6}
\usepackage{array}
\usepackage{here}
\usepackage{amssymb,amsmath}
\usepackage[update,prepend]{epstopdf}

\def\d{{\rm d}}

\begin{document}
\ \
\vspace{0.5mm}
\setcounter{page}{1}

\titlehead{Baltic Astronomy, vol.\,25, 1--8, 2016}

\titleb{Swing amplification and global modes reciprocity \\in models with cusps}

\begin{authorl}
\authorb{E.V. Polyachenko}{1} 
\end{authorl}

\begin{addressl}
\addressb{1}{Institute of Astronomy, Russian Academy of Sciences,\\  Pyatnitskaya 48, Moscow 119017, Russia; epolyach@inasan.ru}
\end{addressl}

\submitb{Received: 2016 June 24; accepted: 2016 xxx xx}

\begin{summary} 
Using 3D N-body simulations we analyse an onset of the bar in cuspy models, and argue that role of swing amplification is twofold. Amplified shot noise due to disc discreteness hampers bar formation, while induced resonance perturbations allow bar amplitude to overcome shots. A bar pattern speed and a growth rate obtained in N-body simulations agree well with global mode analysis. 
 
\end{summary}


\resthead{Swing amplification and global modes reciprocity}{E. V. Polyachenko}

\sectionb{1}{INTRODUCTION}

Among different possibilities for a physical mechanism that is responsible for bar formation in disc galaxies there are two competing especially fiercely. 

The first one is a local approach known now as swing amplification developed in works by Goldreich \& Lynden-Bell (1965), Julian and Toomre (1966), Toomre (1981). Randomly emerging short leading wave packets propagate outward to a forbidden zone boundary near corotation, where they swing into trailing ones. This process is accompanied by amplification of the packets that eventually travel back to the centre. If the packets can freely reflect from the centre, we have an instability mechanism for formation of spiral and barred structures: they are assumed to piece together from the amplified transient waves. However, no one is succeeded yet to calculate earnestly the pattern speed $\Omega_\textrm{p}$ or the growth rate $\omega_\textrm{I}$ of the structures in this frame. 

The second possibility is unstable global modes of axisymmetric stellar discs. Weak bars appear as very open trailing spirals. Here the situation is opposite: at small amplitudes mode's characteristics can be found from linear perturbative analysis using matrix equations (Kalnajs, 1977; Polyachenko 2005) but a physical mechanism hasn't been widely discussed. Lynden-Bell \& Kalnajs (1972) showed that the presence of a trailing wave lowers angular momentum in the inner part of the disc and increases it in the outer parts. The mechanism is basically similar to inverse Landau damping. A steady wave with pattern speed greater than the maximum precession rate of orbits is supported by the central part of the disc (Polyachenko 2004). In the presence of resonances, the wave becomes unstable (Polyachenko 2005). Contrary to the local approach in which an outer Lindblad resonance (OLR) doesn't play any role, for the global modes OLR sometimes is more important than corotation in a sense that transfer of angular momentum to OLR dominates over corotation region (Polyachenko \& Just, 2015). 

The bar forming mechanism should explain appearance of numerical bars in N-body simulations. When using a quiet start technique, which reduces a shot noise (Sellwood \& Athanassoula, 1986), a bar appears as exponentially growing density wave with fixed pattern speed $\Omega_\textrm{p}$ and growth rate $\omega_\textrm{I}$ (e.g., Polyachenko 2013). However, disc discreetness induces the shot noise and secular orbital diffusion (Fouvry et al. 2015). Although not captured by the collisionless Boltzmann equation, they are important ingredients of disc dynamics. In particular, secular changes of the disc DF can lead to instability of initially stable models (Sellwood 2012). 

Using 3D N-body tree-code simulations with $N_\textrm{d} = 6$\,M randomly distributed disc particles in live and rigid cuspy halo/bulge surrounding we analyse an onset of the bar. A key feature that triggers the bar formation in a system with the shot noise is a wave occurring relatively rarely at radii of the order of the disc scale, most likely due to swing amplification of some induced resonance perturbation. We refer to this wave as {\it high wave}. It opposes to {\it low waves}, i.e. more frequent low amplitude waves that are swing amplified shot noise. The high wave increases a bar amplitude to the level invulnerable to the shot noise, where the amplitude starts to grow exponentially fast in accordance with linear perturbation theory based on the pure collisionless Boltzmann equation.  

\sectionb{2}{CALCULATIONS AND RESULTS}

The N-body model we use here is a 3-component model consisting of the stellar disc, S{\'e}rsic cuspy bulge, and NFW dark matter halo with parameters described in Polyachenko et al. (2016ab). 
The bar pattern speed is obtained from an angle of the rotation of the main axes of the inertia ellipsoid. The bar with $\Omega_\textrm{p} \approx 55$ km/s/kpc shows up approximately exponentially. Its growth rate $\omega_\textrm{I}$ is obtained from slopes of the bar strength $B(t)$ or the bar amplitude $A_2/A_0$ (Figure 1), where
\begin{equation}
B(t) = 1 - I_{yy}/I_{xx}\ ,\quad I_{xx}=\sum\limits_{j} \mu_j x^2\ ,\quad A_m(t) = \sum\limits_{j} \mu_j \textrm{e}^{-im\theta_j}\ .
\label{eq:barstr}
\end{equation}
($\mu_j$ and $\theta_j$ are mass and polar angle of star $j$; $j$ spans particles within some fixed radius, e.g., typical radial scale length; expression for $I_{yy}$ is analogous to $I_{xx}$).  In live simulations, the e-folding time $t_\textrm{e}$ is 230 Myr, but in rigid halo/bulge simulations it increases to 530 Myr. 

\begin{figure}[!tH]
\vbox{
\centerline{\psfig{figure=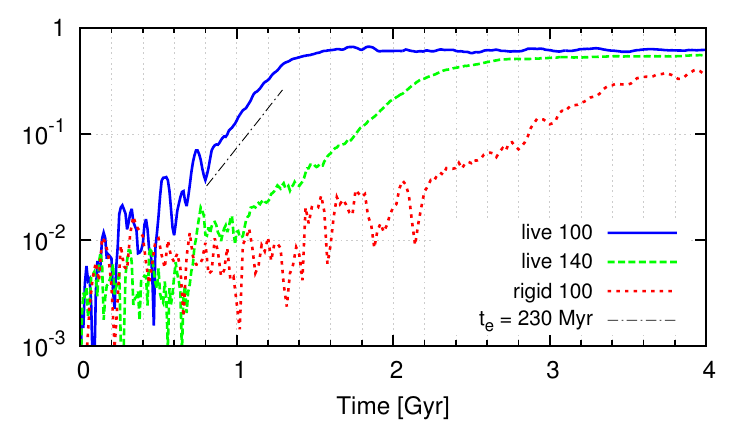,width=67mm,angle=0,clip=}\hspace{1mm}\psfig{figure=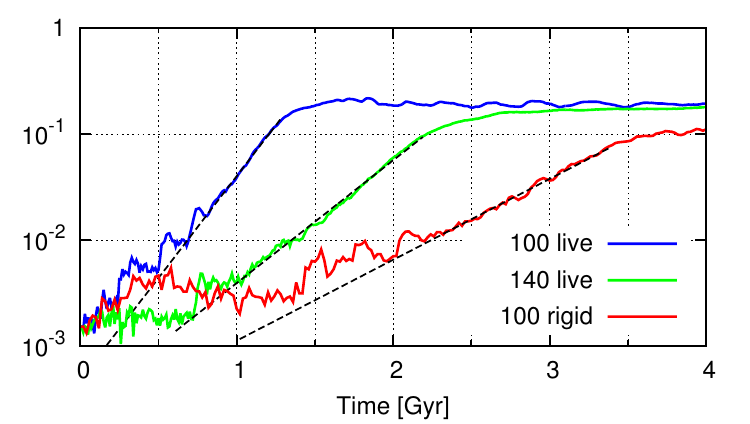,width=67mm,angle=0,clip=}}
\vspace{1mm}
\captionb{1}
{{\it Left panel:} the bar strength for live runs with central radial velocity dispersion $\sigma_{R0} = 100$ and 140 km/s, and rigid halo/bulge run with $\sigma_{R0} = 100$ km/s. The dash-dotted thin line shows a slope corresponding to the exponential growth with $t_\textrm{e}=230$ Myr. {\it Right panel:} the bar amplitude $A_2/A_0$ for the same runs. The dashed black lines show slopes with $t_\textrm{e}=230$, 370, and 530 Myr.}
} 
\end{figure}

Although $B(t)$ and $A_2/A_0$ slopes are nearly the same, the former curves are more irregular. The $B(t)$ curve for 100 km/s central dispersion is raising starting from 0.6 Gyr, the other curves -- from 1 and 2 Gyr.   

Raise of the bar amplitude $A_2/A_0$ is more steady (Figure 1, right panel). Linear fits shown by dashed black lines are calculated using the second half of the raising part and extrapolated to the first half. In all cases we see `jumps' of the colour lines, but they decay on time scales 100 ... 200 Myr. The jumps are absent when the bar amplitude reaches $\sim 1$\,\% of the axisymmetric background.

Bias of the fits from the origin, i.e. {\it lags}, are equal to 0.2, 0.7, 1.2 Gyr, correspondingly. The less unstable model (smaller growth rate) or the number of disc particles $N_\textrm{d}$ is larger, the lag in bar formation is larger. Our tree-code simulations never show it in smaller runs with $N_\textrm{d} = 1.1$\,M. The lag is seen also in works of others (e.g., Dubinski et al. 2009), but it has been neglected.

Rigid halo/bulge models allow matrix calculation of eigen-modes. The obtained pattern speed $\Omega_\textrm{p} \approx 52$ km/s/kpc and the growth rate $\omega_\textrm{I}= 1.8$ ... 1.9 Gyr$^{-1}$ -- both agree well with results of N-body simulations.

Origin of the jumps can be traced by plotting Fourier amplitudes $A_2(R,t)$,
\begin{equation}
A_m(R,t) = {\sum\limits_{j}}' \mu_j \textrm{e}^{-im\theta_j}\ ,
\label{eq:APR}
\end{equation}
where summation is taken over disc particles within a ring $\Delta R$ near $R$. Figure 2, left panel, shows an absolute value of $A_2(R,t)$ for $\sigma_{R0} = 100$ km/s live run. A shot noise exists in all runs (at level $\sim 200$ ... 300), but we show only perturbations with amplitudes higher than 400 by choosing proper colour mapping. A wave with an amplitude 1000--2000 appears at $t\approx 0.2$ Gyr and propagates inward reaching $R=0.58$ kpc at $t\approx 0.47$ Gyr. The arrival is marked by a jump of the blue curve. The subsequent jumps are caused by a series of similar waves born at 4...6 kpc.  

\begin{figure}[!tH]
\vbox{
\centerline{\psfig{figure=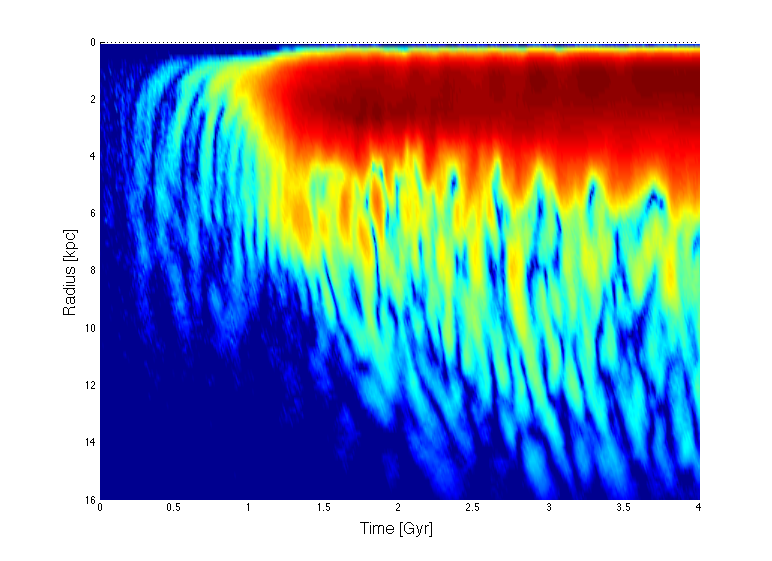,width=67mm,angle=0,clip=}\psfig{figure=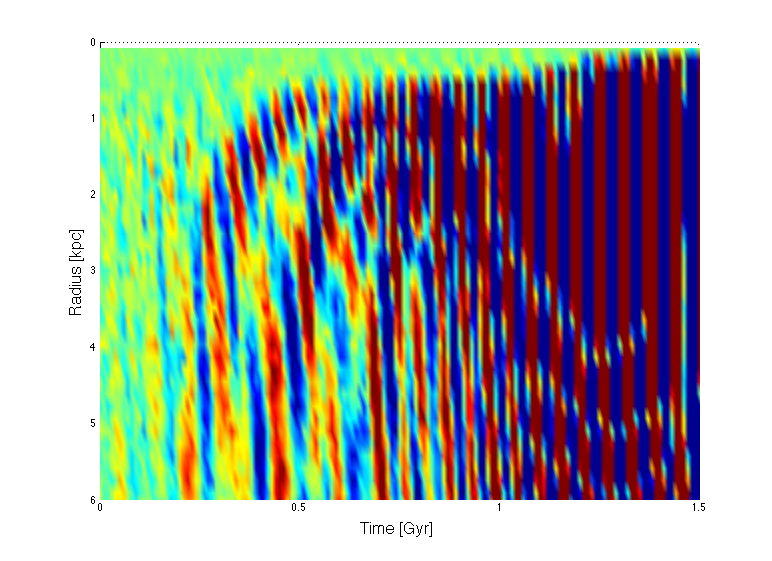,width=67mm,angle=0,clip=}}
\vspace{1mm}
\captionb{2}
{Maps $A_2(R,t)$ for the live run $\sigma_{R0} = 100$ km/s. {\it Left panel:} $\log|A_2(R,t)|$. The colour map is adjusted to show only values larger than 6 (blue). {\it Right panel:} $\Re |A_2(R,t)|$. Green colours show values near zero. A bar-like perturbation is seen as vertical blue and red stripes, while stripes with negative inclination show trailing spirals.}
}
\end{figure}

These waves are usually associated with swing amplified shot noise (e.g., Sellwood 2012). For given parameters $Q$ and $X\equiv k_\textrm{crit}R/m$ favouring swing amplification (Toomre 1981), maximum trailing/leading bias is $\sim 10$ within $2<R<7$ kpc, and shot noise $\sim (R \Sigma_\textrm{d})^{1/2}$ is large. However, this bias is measured for the maximum of the trailing amplitude; the ratio for amplitudes far from amplification zone is much smaller. Besides, the given factor is for a full swing; e.g., for half-swing it is less than $\sim 3$. The live halo increases this value, but it is hardly possible to obtain values more than 1000. Typical values for such swing amplified noise well seen in Figure 3 (right) are 200-400. We denote these waves as {\it low waves}. 

Another possibility is swing amplified induced perturbations. Such perturbations are possible as responses to the gravitation potential of the emerging weak bar in the centre (Polyachenko 2002) or spiral modes of the disc. In favour of this hypothesis, one can observe an increasing activity in formation of the high amplitude waves ($>1000$) at $t>0.6$ Gyr in Figure 2, and $t>1$ and 1.5 Gyr at Figure 3. We denote these waves as {\it high waves}. 

In Figure 2, right panel, real part of $A_2(R,t)$ is presented in a smaller range of time and radius. Before $t < 0.2$ Gyr we can see ordered perturbations in form of trailing waves with amplitudes 200--400. Two high waves appear during first 0.6 Gyr, which disappear after $t\gtrsim 0.8$ Gyr in the bar area (vertical stripes).

Less unstable models are given in Figure 3. The left panel shows a map for a live run with increased radial velocity dispersion. The first wave appears only at $t\approx 0.74$ Gyr and causes the first jump in the bar amplitude curve (Figure 1, green line). Note that no growth of the amplitude is seen before the jump, so the wave triggers the bar formation. Growth of the bar amplitude becomes approximately exponential after the second wave that forms at $\sim 1$ Gyr. 

A rigid run with $\sigma_{R0} = 100$ km/s (right panel) presents even more irregular bar formation. The first wave appears at nearly the same time as in the live run (Figure 2), but it is not supported by the subsequent waves. This leads to a gradual decay of the bi-symmetric perturbation from 0.5 to 1.5 Gyr (Figure 1, red line) until the high waves do not restart process of bar formation.

\begin{figure}[!tH]
\vbox{
\centerline{\psfig{figure=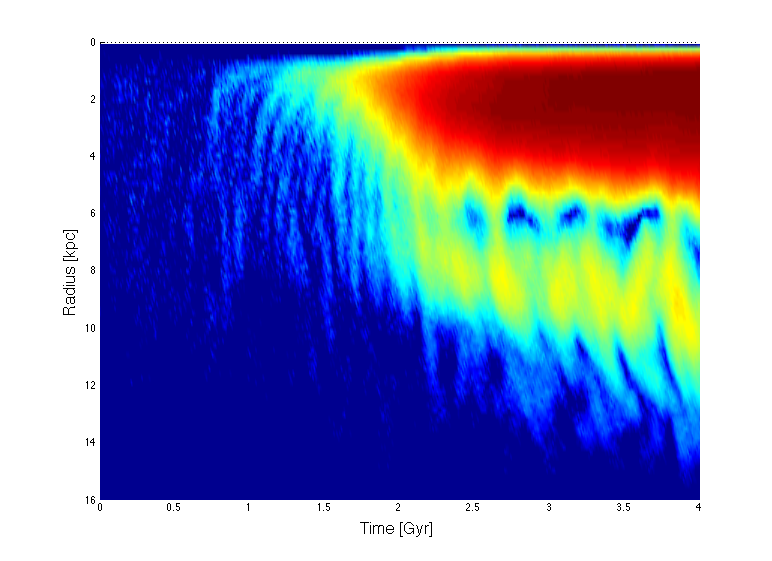,width=67mm,angle=0,clip=}\psfig{figure=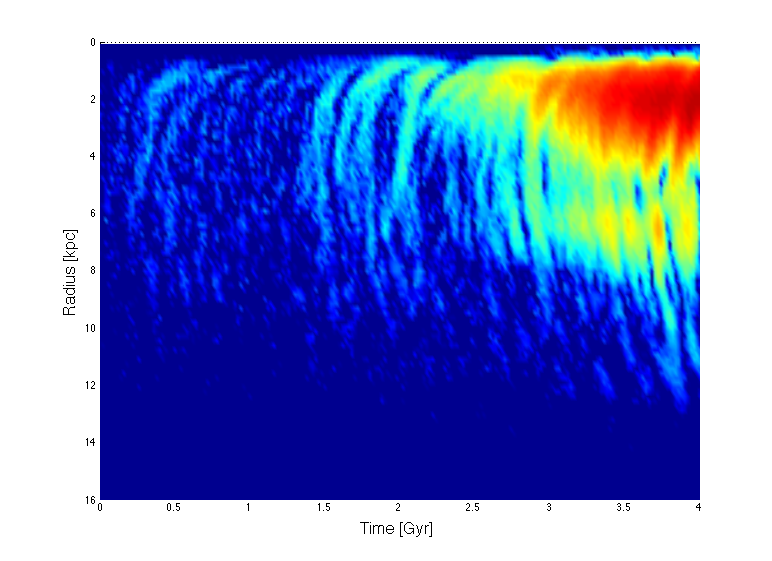,width=67mm,angle=0,clip=}}
\vspace{1mm}
\captionb{3}
{Same as in the left panel of Figure 2, for the live run $\sigma_{R0} = 140$ km/s and the rigid run $\sigma_{R0} = 100$ km/s.}
}
\end{figure}
The time Fourier transform of $A_m$:
\begin{equation}
S_R(R,\Omega_\textrm{p}, T) = \int\limits_{T-\Delta T}^{T+\Delta T} \d t \, A_m(R,t) H(t-T) \textrm{e}^{im\Omega_\textrm{p} t} \ ,
\label{eq:Sr}
\end{equation}
with filter function $H$, produces a spectral map
\begin{equation}
M_R(R,\Omega_\textrm{p}, T) = \ln \frac{|S_R|^2}{2\pi}
\label{eq:M}
\end{equation}
providing locations of coherent structures in $(R,\Omega_\textrm{p}$)-plane. 

Figure 4 shows $M_R$ of bi-symmetric ($m=2$) harmonics. In the earliest panel we see 3 distinct maxima corresponding to modes with different pattern speeds. The top maximum corresponds to a bar mode, which is in fact an open spiral with average pitch angle $\sim 60^\circ$. 
Two other maxima, localised between 1.5 and 6 kpc and 2 and 8 kpc, can be spiral modes that seed the incoming waves. Indeed, these maxima look like modes, remaining fixed in the $(R,\Omega_\textrm{p}$)-plane almost until the bar is formed. Interference of these modes can locally form high enough over-density, that is eventually amplified by the swing mechanism.
The same three maxima are seen in Figure 5 calculated for the rigid halo/bulge run. Note that bar mode is seen already in the first (upper left) panels of Figures 4 and 5.

\begin{figure}[!tH]
\vbox{
\centerline{\psfig{figure=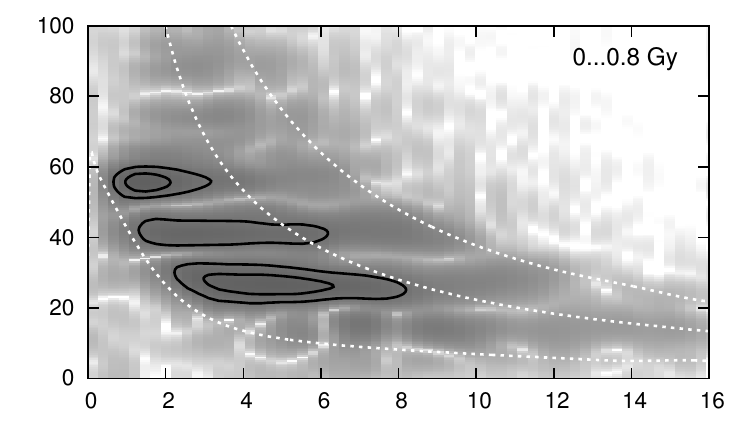,width=67mm,angle=0,clip=}\psfig{figure=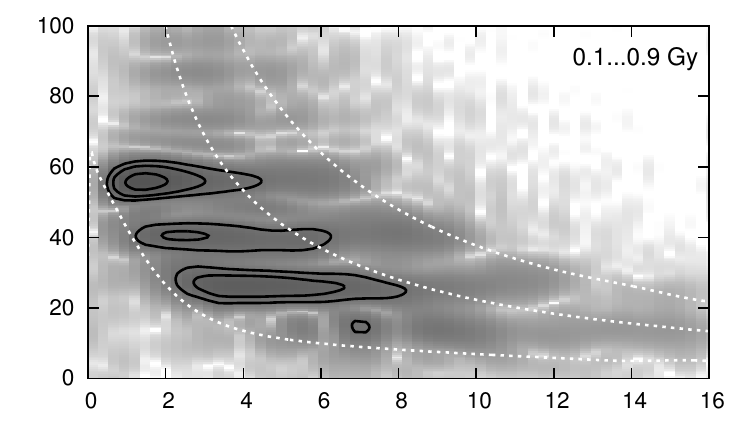,width=67mm,angle=0,clip=}}
\centerline{\psfig{figure=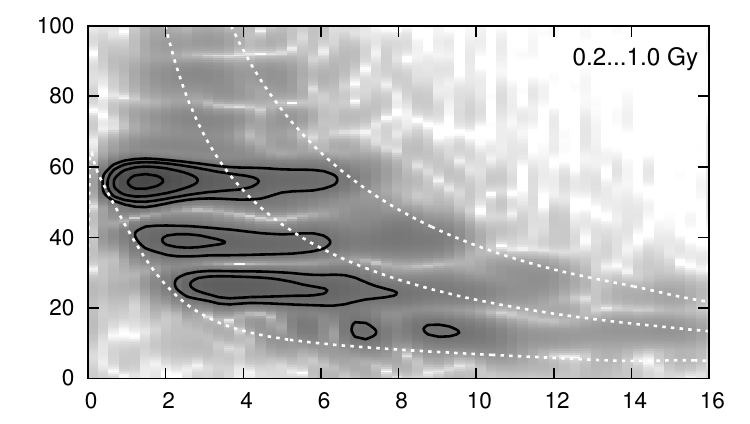,width=67mm,angle=0,clip=}\psfig{figure=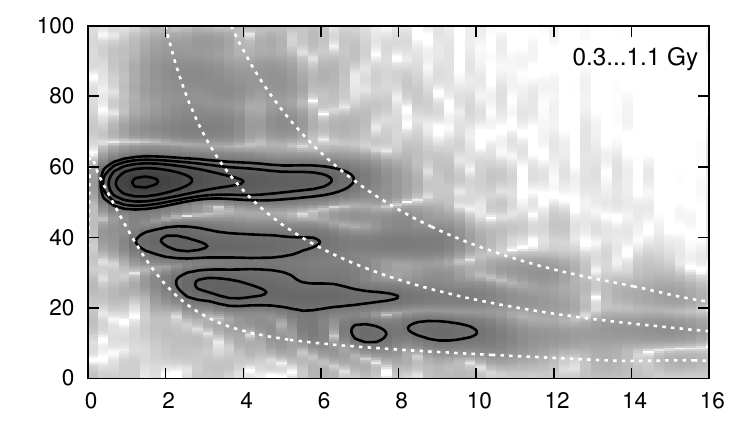,width=67mm,angle=0,clip=}}
\centerline{\psfig{figure=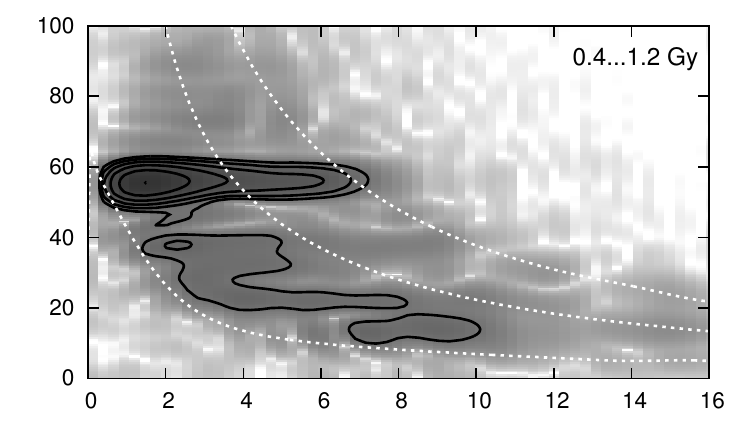,width=67mm,angle=0,clip=}\psfig{figure=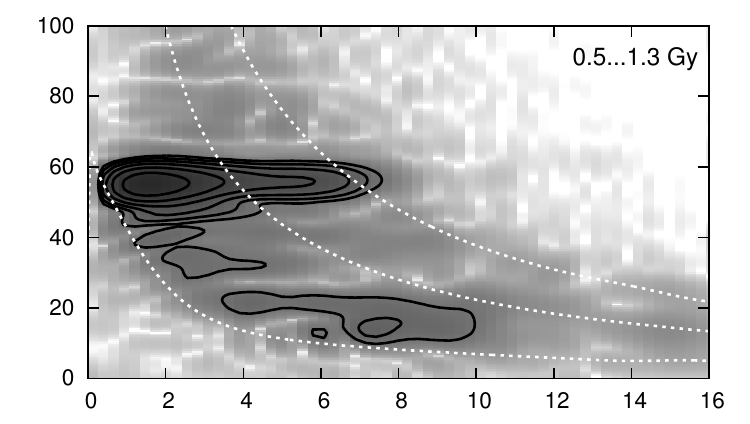,width=67mm,angle=0,clip=}}
\vspace{1mm}
\captionb{4}
{Spectral maps $M_R$ of bi-symmetric harmonics for different time windows (given in the upper right corners). Contour levels are evenly spaced in log scale and keep fixed through the panels. Angular velocity profile $\Omega(R)$, and $\Omega(R) \pm\kappa/2$ indicating positions of Lindblad resonances, are shown by dashed white lines. Radius along x-axis is in kpc, frequency along y-axis is in km/s/kpc.}
}
\end{figure}

\begin{figure}[!tH]
\vbox{
\centerline{\psfig{figure=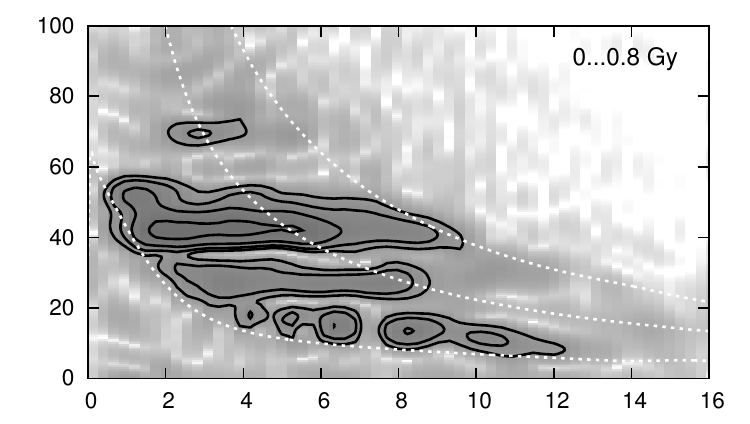,width=67mm,angle=0,clip=}\psfig{figure=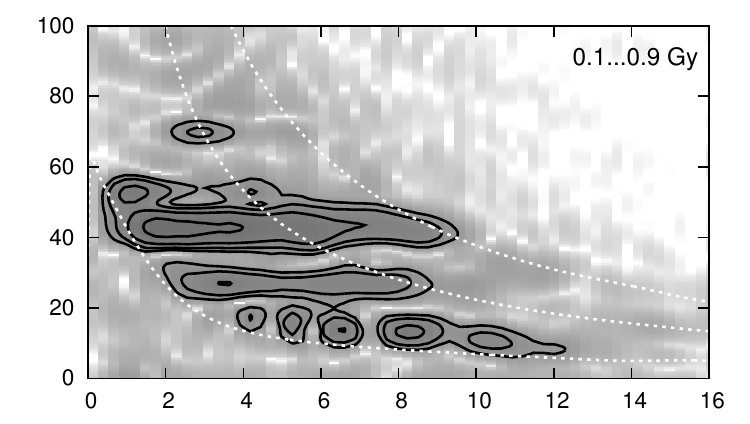,width=67mm,angle=0,clip=}}
\centerline{\psfig{figure=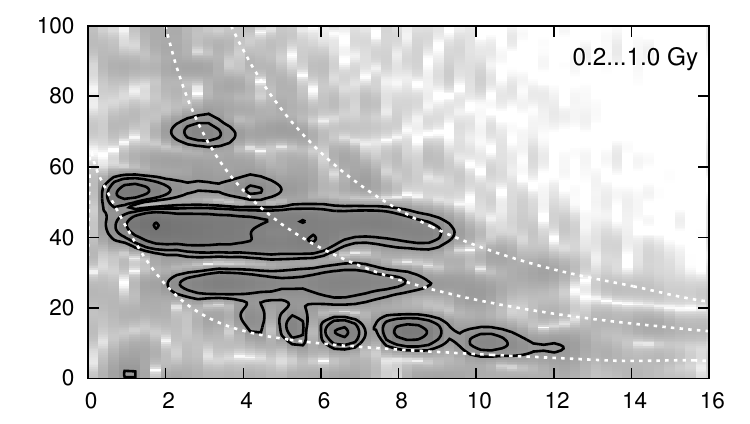,width=67mm,angle=0,clip=}\psfig{figure=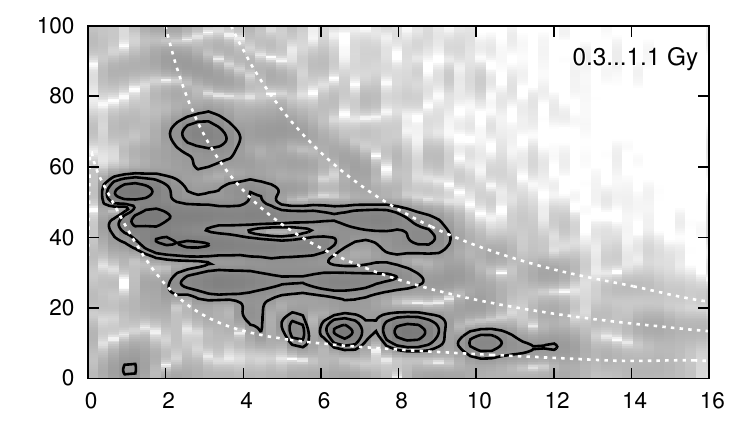,width=67mm,angle=0,clip=}}
\centerline{\psfig{figure=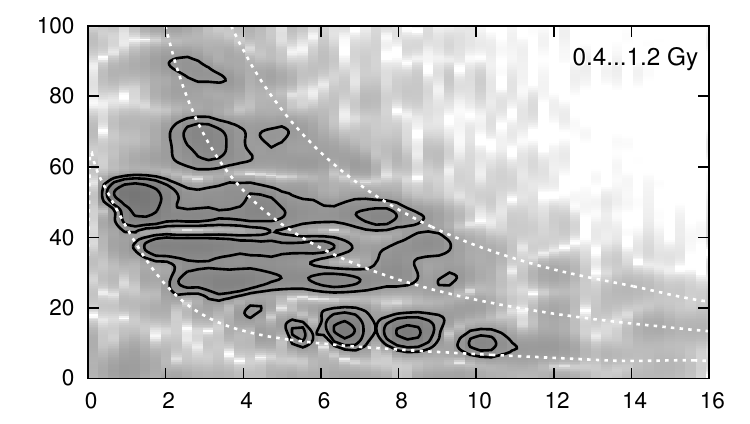,width=67mm,angle=0,clip=}\psfig{figure=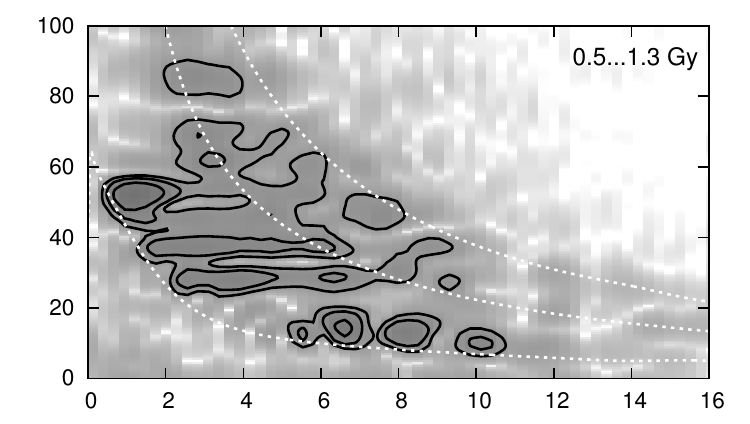,width=67mm,angle=0,clip=}}
\vspace{1mm}
\captionb{5}
{Same as in Figure 4, for the rigid run.}
}
\end{figure}

\sectionb{3}{DISCUSSION AND CONCLUSIONS}

A real N-body system is a complex interplay of instability well described by collisionless Boltzmann equation, and the shot noise effects. Here we list our observations based on analysis of several N-body simulations:
\begin{itemize}
\item there is a shot noise background in the stellar disc;
\item at some moment, a wave appears at $\sim 4$...6 kpc and propagates to the centre; 
\item plausibly, the swing mechanism amplifies a seed perturbation into the wave;
\item spectral maps show several modes, including bar mode: these modes can provide the seed for the wave;
\item arrival of the wave in the centre triggers bar formation;
\item since the wave appears randomly, there is a random lag in bar formation; 
\item the wave causes a jump in the bar amplitude, which relaxes gradually; 
\item other waves appear randomly and cause similar jumps until it becomes about 1\,\% of the axisymmetric background; afterwards the bar amplitude grows exponentially;
\item the exponential growth rate agrees well with predictions based on the collisionless Boltzmann equation;
\item the larger the growth rate and smaller the number of disc particles, the smaller the lag. 
\end{itemize}

These observations fit to a following scenario. An emerging bar mode due to the global mode instability is kicked by two kinds of incident waves. Both kinds of waves are due to swing mechanism. Low amplitude waves that come from the shot noise are frequent and they are not in phase with the bar, so they hamper bar formation. High amplitude waves that induced by interference of modes are rare, but they can increase the bar amplitude. When the latter becomes high enough, it starts to grow uniformly due to usual global mode instability.

V.I. Korchagin reported (private communication) that introduction of annulus in the region within corotation wiping out all non-axisymmetric perturbations leads to faster bar formation. Our hypothesis explains this effect. Indeed, when all incoming waves (both low and high) are wiped out, bar forms as usual unstable mode with no obstruction.    

\thanks{The runs was done on the {\tt MilkyWay} supercomputer, funded by the Deutsche Forschungsgemeinschaft (DFG) through Collaborative Research Centre 
(SFB 881) ``The Milky Way System'', hosted and co-funded by the J\"ulich Supercomputing Center (JSC), and GPU cluster {\tt kepler}, funded under the grants 
I/80 041-043 and I/81 396 of the Volkswagen Foundation and grants 823.219-439/30 and /36 of the Ministry of Science, Research and the Arts of 
Baden-W\"urttemberg, Germany. This work was supported by Sonderforschungsbereich SFB 881 ``The Milky Way System'' (subproject A6) of the German Research Foundation (DFG), by Volkswagen Foundation under the Trilateral Partnerships grant No. 90411, by Russian Basic Research Foundation, grants 15-52-12387, 16-02-00649, and by Basic Research Program OFN-15 `The active processes in galactic and extragalactic objects' of Department of Physical Sciences of  RAS. }

\References


\refb Dubinski J., Berentzen I., Shlosman I. 2009, ApJ, 697, 293

\refb Goldreich P., Lynden-Bell D. 1965,  MNRAS, 130, 125

\refb Julian W. H., Toomre A. 1966, ApJ, 146, 810

\refb Kalnajs A. 1977, ApJ, 212, 637

\refb Lynden-Bell D., Kalnajs A., 1972, MNRAS, 157, 1




\refb Polyachenko E. V. 2002, MNRAS, 330, 105 

\refb Polyachenko E. V. 2004, MNRAS, 348, 345

\refb Polyachenko E. V. 2005, MNRAS, 357, 559

\refb Polyachenko E. V. 2013, Astron. Lett, 39, 72

\refb Polyachenko E. V., Berczik P., Just A. 2016a, Submitted to MNRAS

\refb Polyachenko E. V., Berczik P., Just A. 2016b, Submitted to Baltic Astronomy

\refb Polyachenko E. V., Just A. 2015, MNRAS, 446, 1203

\refb Sellwood J.A. 2012, ApJ, 751, 44

\refb Toomre A., {\it Structure and evolution of normal galaxies} (Ed. S.M. Fall and D. Lynden-Bell, Cambridge Univ. Press., 1981), p.111.


\end{document}